\documentclass[prx,  
 reprint,
superscriptaddress, 
longbibliography, 
 amsmath,amssymb,
 aps,
]{revtex4-2}     

\usepackage[authormarkup=superscript,]{changes} 

\definecolor{addedcolor}{RGB}{204, 0, 0} 
\definecolor{deletedcolor}{RGB}{204, 0, 0} 
 
\setaddedmarkup{\textcolor{addedcolor}{#1}} 
\setdeletedmarkup{\textcolor{deletedcolor}{\sout{#1}}} 
 
\usepackage{graphicx}
\usepackage{dcolumn}
\usepackage{bm}
\usepackage{float}  
\usepackage[none]{hyphenat}  

\usepackage[utf8]{inputenc}  
\usepackage[T1]{fontenc} 

\usepackage{comment} 
 
\usepackage[colorlinks=true,linkcolor=blue,allcolors=blue]{hyperref}

\newcommand{\aveNa}{\langle n_a \rangle}       

\newcommand{\fr}{r_i}

\newcommand{\aveS}{\langle S \rangle}  
\newcommand{\aveT}{\langle D \rangle}        
\newcommand{\aveX}{\langle X \rangle}    
\newcommand{\quietFraction}{\delta_\textrm{aval}}      
\newcommand{\aveTau}{\langle \tau \rangle}

\newcommand{\aveR}{ \rho }    

\newcommand{\pin}{ p_\textrm{in} }  
\newcommand{\durec}{ \delta u_\textrm{rec} }

\usepackage{xcolor}    

\newcommand{\mycomment}[1]{}

\begin{document}

\title{ Allometric scaling of brain activity explained by avalanche criticality } 

\author{Tiago S. A. N. Sim\~{o}es}
\affiliation{University of Campania “Luigi Vanvitelli”, Department of Mathematics and Physics, Caserta, Viale Lincoln, 5, 81100, Italy} 

\author{José S. Andrade Jr.}  
\affiliation{Universidade Federal do Cear\'a, Departamento de F\'isica, Fortaleza, Cear\'a, 60451-970, Brazil} 

\author{Hans J. Herrmann}  
\affiliation{Universidade Federal do Cear\'a, Departamento de F\'isica, Fortaleza, Cear\'a, 60451-970, Brazil} 
\affiliation{ESPCI, PMMH, Paris, 7 quai St. Bernard, 75005, France} 

\author{Stefano Zapperi} 
\affiliation{Center for Complexity and Biosystems, Department of Physics, University of Milan, via Celoria 16, 20133 Milano, Italy} 
\affiliation{CNR - Consiglio Nazionale delle Ricerche, Istituto di Chimica della Materia Condensata e di Tecnologie per l'Energia, Via R. Cozzi 53, 20125 Milano, Italy}   

\author{Lucilla de Arcangelis}    
\affiliation{University of Campania “Luigi Vanvitelli”, Department of Mathematics and Physics, Caserta, Viale Lincoln, 5, 81100, Italy}   

\begin{abstract}  
Allometric scaling laws, such as Kleiber’s law for metabolic rate, highlight how efficiency emerges with size across living systems. The brain, with its characteristic sublinear scaling of activity, has long posed a puzzle: why do larger brains operate with disproportionately lower firing rates? Here we show that this economy of scale is a universal outcome of avalanche dynamics. We derive analytical scaling laws directly from avalanche statistics, establishing that any system governed by critical avalanches must exhibit sublinear activity–size relations. This theoretical prediction is then verified in integrate-and-fire neuronal networks at criticality and in classical self-organized criticality models, demonstrating that the effect is not model-specific but generic. The predicted exponents align with experimental observations across mammal species, bridging dynamical criticality with the allometry of brain metabolism. Our results reveal avalanche criticality as a fundamental mechanism underlying Kleiber-like scaling in the brain.    
\end{abstract}  

\maketitle

\section{Introduction} 

Why do larger brains fire more efficiently? Scaling laws have long revealed how living systems achieve remarkable economies of scale, from metabolism to morphology. In biology, allometry denotes systematic departures from isometry, where traits grow with body size according to power laws \cite{huxley_terminology_1936,shingleton_allometry_2010}. Among the most celebrated examples is Kleiber’s law, describing how metabolic rate scales with body mass as $r_m \sim M^{3/4}$ \cite{kleiber_body_1932,kleiber_body_1947,burger_allometry_2019, west_general_1997,banavar_general_2010}. The brain follows its own distinct rules: brain mass scales sublinearly with body mass, $m_b \sim M^{3/4}$, unlike most organs that grow closer to isometry \cite{burger_allometry_2019,peters_ecological_1983}. Other brain traits show equally systematic patterns, from the metabolic rate scaling with brain volume $r_m \sim V_b^{5/6}$ \cite{karbowski_global_2007,herculano-houzel_scaling_2011,herculano-houzel_remarkable_2012} to the near-isometric relation between brain mass and neuron number in rodents, which contrasts with the superlinear scaling in primates \cite{herculano-houzel_human_2009}. Cortical folding, too, varies by lineage, scaling shallowly in rodents but more steeply in primates \cite{ventura-antunes_different_2013}. These regularities suggest that brain function is governed by an economy of scale, yet the dynamical origin of this allometry remains unresolved \cite{shingleton_allometry_2010,burger_allometry_2019}.
 
\begin{figure*}[!htb]  
    \centering
    \includegraphics[width=\textwidth]{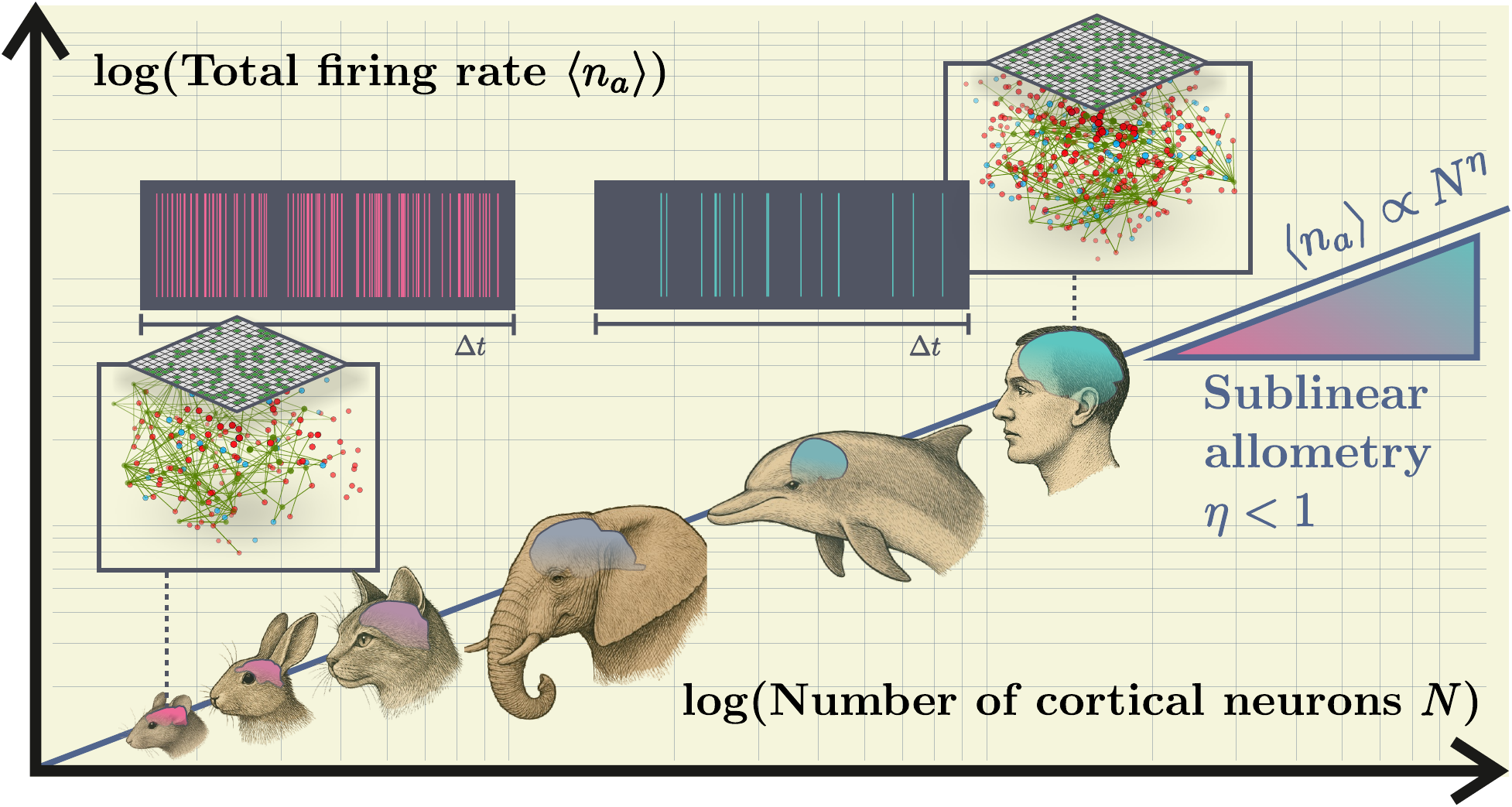}
    \caption{Illustration of sublinear allometry in a system exhibiting avalanche dynamics: the neuronal cortex. As the number of cortical neurons $N$ increases, the total firing rate $\langle n_a \rangle$ is expected to scale sublinearly with $N$ in an economy of scale regime, as $\langle n_a \rangle \propto N^{\eta}$ with $\eta < 1$, due to the spatio-temporal organization of cortical activity into neuronal avalanches. This sublinear scaling implies that individual neurons fire, on average, more frequently in smaller brains (e.g., mouse, red spikes) than in larger ones (e.g., human, blue spikes).}
    \label{fig:allometry-figure}
\end{figure*} 

\begin{figure}[!htb] 
    \centering
    \includegraphics[width=\columnwidth]{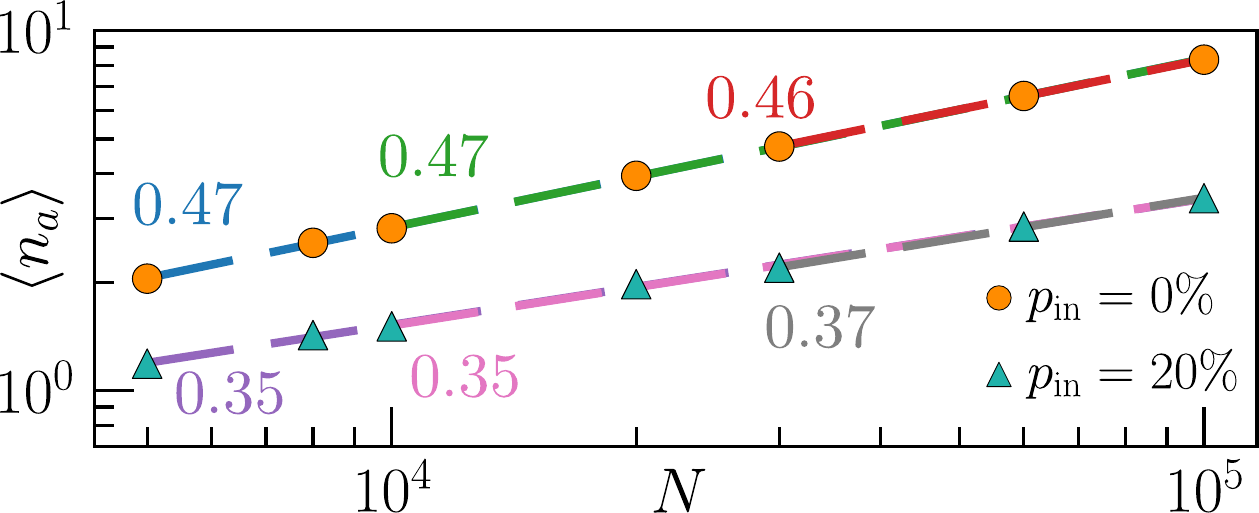}
    \caption{Numerical results for the average number of firing neurons $\aveNa$ during each timestep as a function of $N$, for fully-excitatory networks (orange symbols) and networks with $\pin = 20 \%$ (cyan symbols). The coloured dashed lines are power-law least-square fits $\aveNa \propto N^{\eta}$.}
    \label{fig:if-allometry}
\end{figure}   

In parallel, mounting evidence shows that neural activity performs close to criticality. Neuronal avalanches—spatiotemporal cascades of firings with power-law distributed sizes and durations—were first observed in rat cortex \cite{beggs_neuronal_2003} and subsequently across other species, including humans \cite{petermann_spontaneous_2009,mora_dynamical_2015,hahn_spontaneous_2017,ponce-alvarez_whole-brain_2018,shriki_neuronal_2013}. These cascades are hallmarks of self-organized criticality (SOC) \cite{bak_self-organized_1987,zapperi_crackling_2022,jensen_self-organized_1998}, first introduced in the Bak–Tang–Wiesenfeld sandpile model \cite{bak_self-organized_1987} and formalized through branching processes \cite{zapperi_self-organized_1995,vespignani_order_1997}. SOC provides a natural explanation for how complex systems self-tune to critical states without fine control.   
Here, we propose that the sublinear allometry of brain activity is not merely an echo of metabolic scaling, but a universal outcome of avalanche criticality, as illustrated in Fig.\ \ref{fig:allometry-figure}.  
By deriving scaling relations directly from avalanche statistics, we establish that any avalanching system can exhibit sublinear activity–size relations. We then confirm these predictions in neuronal networks at criticality and in classical SOC models, demonstrating that the phenomenon is not model-specific but generic. Finally, by comparing with experimental data in mammals \cite{karbowski_thermodynamic_2009,herculano-houzel_scaling_2011,karbowski_scaling_2011,lennie_cost_2003,yu_evaluating_2017}, we argue that the economy of scale of the brain reflects a universal dynamical law of avalanching systems.    

\begin{figure*}[!htb]              
    \centering
    \includegraphics[width=\textwidth]{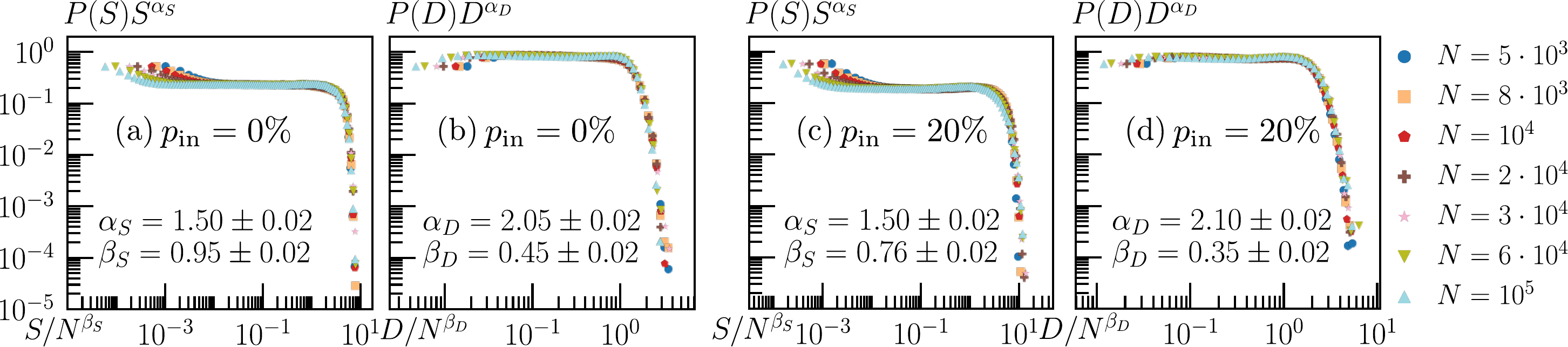}     
    \caption{
    Collapse of the numerical distributions of avalanche sizes $S$ (a and c) and durations $D$ (b and d), for different sizes $N \in [ 5 \cdot 10^3 , 10^5 ]$, onto a universal curve $\tilde{P}(S) = S^{\alpha_S} \mathcal{F}(S/N^{\beta_S})$ and $\tilde{P}(D) = D^{\alpha_D} \mathcal{F}(D/N^{\beta_D})$, respectively, considering IF networks with two different fractions of inhibitory neurons $\pin = 0 \%$ (a-b) and $\pin = 20 \%$ (c-d).}
    \label{fig:if-avalanches}         
\end{figure*}          
         
\section{Results} 
\subsection{Sublinear allometry in integrate-and-fire networks}

To analyse how neuronal activity scales with system size under controlled conditions, we consider an integrate-and-fire (IF) model implemented on a scale-free, directed neural network of $N \in [5 \cdot 10^3 , 10^{5}]$ neurons, with short- and log-term plasticity \cite{michiels_van_kessenich_critical_2018}. 
A neuron can be either excitatory or inhibitory, and fires whenever its voltage reaches a threshold $v_c = 1$. We consider both fully-excitatory networks, $\pin = 0 \%$, and networks with a fraction $\pin = 20 \%$ of inhibitory neurons. When the voltage of all neurons is below the threshold, external input is introduced by adding a small voltage $\delta v = 0.1$ to a randomly selected neuron at each timestep, independently of $N$. This ensures a slow external driving \cite{vespignani_order_1997,laurson_effect_2009}, an important constraint to observe avalanche-like behaviour \cite{vespignani_absorbing-state_2000}. The model includes a parameter $\durec$ that, when properly tuned for each system size $N$ and fraction $\pin$, sets the system in the critical state \cite{michiels_van_kessenich_critical_2018} (see Section \ref{section:Methods} for details of the implementation).        
We then study the average number of firing neurons $\aveNa$ per timestep $\Delta t$ as a function of the system size $N$,  
\begin{equation}
    \aveNa = \Delta t \sum_{i}^{N}r_i \equiv \sum_{i}^{N}r_i \textrm{ ,}  
\end{equation}    
where we set $\Delta t = 1$ without loss of generality. The quantity $\fr$ is the firing rate of neuron $i \in [1,N]$, defined as $ r_{i} = ( 1 / T ) \sum_{t=1}^{ T } \sigma_{i}(t)$, with $\sigma_{i} (t) = 1$ if $v_{i} \geq v_c$ during the $t$-th timestep, $\sigma_{i} (t) = 0$ otherwise, and $T$ is the total number of timesteps.  

In Fig.\ \ref{fig:if-allometry} we show the numerical results for $\aveNa$ as a function of $N$, for the IF model considering fully-excitatory networks and networks with $\pin = 20 \%$, in the critical state. 
In that same figure, we also fit the numerical data with a power-law $\aveNa \propto N^{\eta}$ (coloured dashed lines) by considering consecutively less data points for small system sizes, to take into account the weak curvature in the data and study how the exponent $\eta$ converges for large $N$. 
In both cases, we observe sublinear allometric scaling, with $ \eta = 0.46 \pm 0.04 $ for fully-excitatory networks and an even slower-than-linear scaling with $\eta = 0.37 \pm 0.04$ for networks with $\pin = 20 \%$, considering $N \in [3 \cdot 10^4, 10^5]$ (red and gray dashed lines). 
This scaling relation implies that neuronal activity operates according to an ``economy of scale'' regime \cite{bettencourt_growth_2007,oliveira_large_2014,melo_statistical_2014}, whereby larger biological neuronal networks exhibit disproportionately lower activity rates.     
 
In Fig.\ \ref{fig:if-avalanches} we plot the collapse of the distributions $P(S)$ and $P(D)$ of avalanche sizes $S$ and durations $D$, respectively, for several $N$ and both $\pin = 0 \%$ and $\pin = 20 \% $ (see supplementary information for the non-collapsed data).        
Both distributions follow a universal scaling relation, corresponding to $\tilde{P}(S) = S^{\alpha_S} \mathcal{F}(S/N^{\beta_S})$ for the size distributions and $\tilde{P}(D) = D^{\alpha_D} \mathcal{F}(D/N^{\beta_D})$ for the duration distributions. The values of $\alpha_S \approx 3/2$ and $\alpha_D \approx 2$ are characteristic of the universality class of the mean-field self-organized branching process \cite{zapperi_self-organized_1995}, and are consistent with experimental observations \cite{beggs_neuronal_2003,shriki_neuronal_2013}.

\subsection{Scaling of \texorpdfstring{$\boldsymbol{\langle n_a \rangle}$}{<n_a>} with \texorpdfstring{$\boldsymbol{N}$}{N}: Analytical derivation}  

At this point, we propose that the allometric scaling shown in Fig.\ \ref{fig:if-allometry} is a consequence of avalanche dynamics and therefore depends on the finite-size scaling behaviour of the distributions in Fig.\ \ref{fig:if-avalanches}. To demonstrate this statement, we generalize $\aveNa$ to represent the average number of active sites per timestep in a generic model with avalanche dynamics, where $r_i$ now denotes the average activation rate of site $i \in [ 1 , N ]$. 
First notice that $\aveNa$ is equivalent to the ratio $\aveS / \aveT$ multiplied by the fraction of time $ \quietFraction \leq 1$ involved in avalanching processes, where $\aveS$ and $\aveT$ are the average avalanche size and duration, respectively. The fraction $\quietFraction$ can be estimated by considering the average quiet time $\aveTau$ between avalanches, $ \quietFraction = \aveT / ( \aveT + \aveTau ) $, therefore   
\begin{equation}  
	\label{eq:ave_Na_formula}       
	\aveNa = \frac{ \aveS }{ \aveT + \aveTau } \textrm{ .}   
\end{equation}         
Since a particular site may activate several times during an avalanche, we note that, in Eq.\ (\ref{eq:ave_Na_formula}), $S$ represents the total number of activation events during an avalanche, as opposed to the number $S_d \leq S$ of distinct activated sites.  
       
In the critical state, according to the finite-size scaling assumption \cite{chessa_critical_1999}, $S$ and $D$ follow a power-law distribution with a cut-off that scales with the system size,   
\begin{equation} \label{eq:pX}    
    P(X) = A_X f(X) X^{-\alpha_X} g(X/X_c) \textrm{, for $X \in [1, \infty[$, }  
\end{equation} 
where $X \in \{ S, D \}$, $A_X$ is the normalization constant, and $f(X)$ is a correction factor which accounts for possible deviations from the power-law behaviour at small $X$, with $f(X) \approx 1$ when $X \gg 1$. $g(X/X_c)$ is a rapidly decaying function for $X > X_c$ modelling the cut-off, where $X_c \propto N^{\beta_X}$, and $g(X/X_c) \approx 1$ for $ X < X_c$. Given these definitions, the scaling of the average $\aveX = \int_{1}^{\infty} X P(X) dX$ can be estimated in the large $N$ limit (see supplementary information),
\begin{equation} \label{eq:aveXLargeN}   
    \lim_{N\to\infty} \langle X \rangle \propto N^{\beta_X \cdot (2 - \alpha_X)} \textrm{ , for $ 1 < \alpha_X < 2 $.}      
\end{equation}   
Equation (\ref{eq:aveXLargeN}) is consistent with previous analytical results \cite{chessa_universality_1999,huynh_abelian_2011}.
For $\alpha_X \leq 1$ or $\alpha_X \geq 2$, corrections to Eq.\ (\ref{eq:aveXLargeN}) are needed. In particular, if $\alpha_X > 2$, $\lim_{N\to\infty} \aveX$ is expected to converge to a constant, while for $\alpha_X = 1$ or $\alpha_X = 2$, modifications to the exponent are needed and an additional logarithmic scaling with $N$ is also present (see supplementary information). 
  
\begin{figure}[!htb]        
    \centering
    \includegraphics[width=\columnwidth]{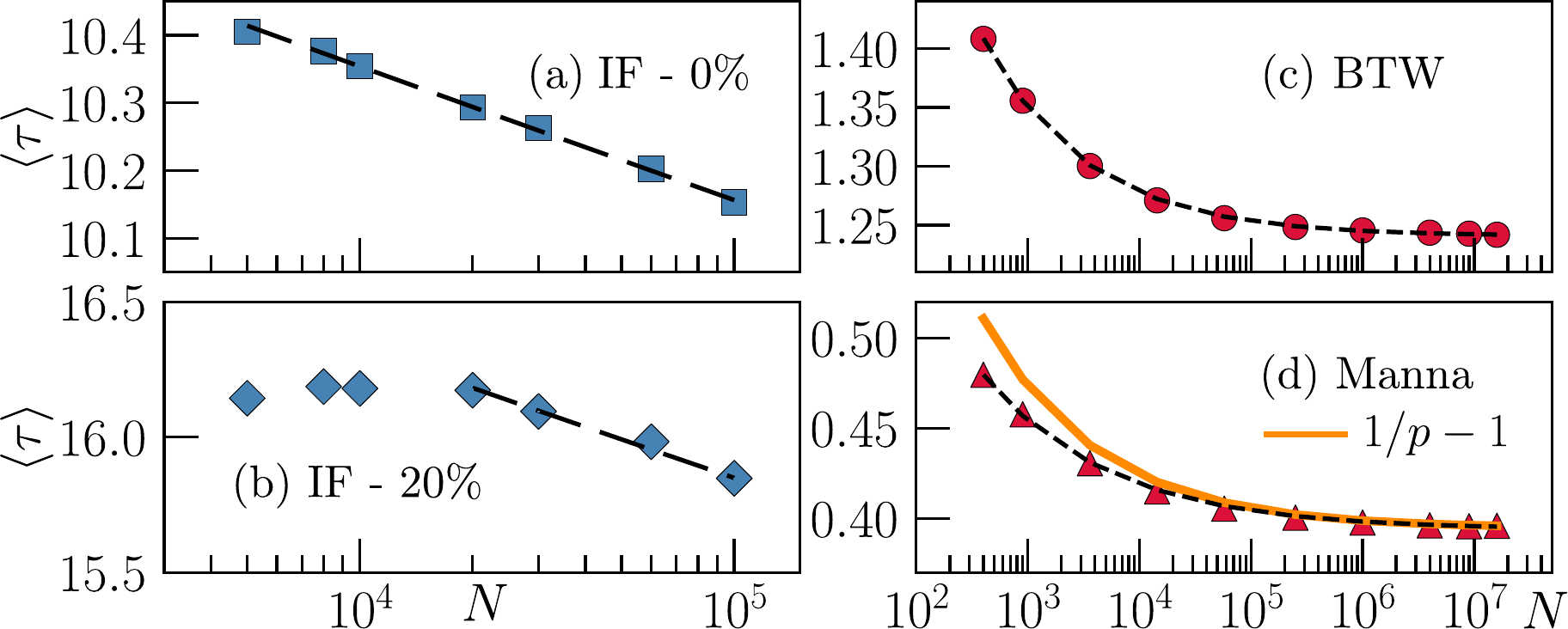}
    \caption{Numerical results for the average quiet time $\aveTau$ as a function of the system size $N$ for the IF model with $ \pin = 0 \% $ (a) and $ \pin = 20 \% $ (b) inhibitory neurons, as well as for the BTW (c) and Manna (d) sandpile models. For all models, at least for large $N$, the data are well fitted by $\aveTau = a N^{b} + c$ (black dashed lines), where $\{a,b,c\}$ are the fitting parameters. For all cases, a non-linear least squares fit gives $b < 0$, meaning $\lim_{N\to\infty} \aveTau$ tends to a constant: $ c = 5 \pm 5$ (IF, $\pin= 0\%$), $ c = 7.5 \pm 7.5 $ (IF, $\pin = 20 \%$); $c=1.24 \pm 0.01$ (BTW); $c = 0.39 \pm 0.01$ (Manna). In (d) we also plot the quantity $1/p - 1$ (orange line) for several $N$, where $p$ is the time-averaged density of occupied sites (sites with one grain). As explained in the supplementary information, $1/p - 1$ should equal $\aveTau$ in the Manna model, in the limit $N \rightarrow \infty$.}
    \label{fig:aveTauNew}
\end{figure}

By joining Eqs.\ (\ref{eq:ave_Na_formula}) and (\ref{eq:aveXLargeN}), we have   
\begin{equation} \label{eq:aveNaLargeN}                                 
    \lim_{N\to\infty} \aveNa \propto \frac{ N ^ { \beta_S \cdot (2 - \alpha_S) } }{ c_D N ^ { \beta_D \cdot (2 - \alpha_D) } + \aveTau  }  \textrm{ , for $ 1 < \alpha_S, \alpha_D < 2$,}   
\end{equation} 
where $c_D$ is a fitting parameter, corresponding to the amplitude of the scaling of $\lim_{N\to\infty} \aveT$. Eq.\ (\ref{eq:aveNaLargeN}) predicts that, in general, depending on the behaviour of $\aveTau$, the average number of active sites $\aveNa$ scales non-linearly with $N$.  
In Fig.\ \ref{fig:aveTauNew}(a-b) we present the numerical results for the average time $\aveTau$ for the IF model, as well as a fitting with the function $\aveTau = aN^{b} + c$, with $\{a,b,c\}$ being fitting parameters (dashed line). 
 
We notice that $ \lim_{N\to\infty} \aveTau $ tends to a constant for both cases $\pin=0 \%$ and $\pin = 20 \% $, thus, for large $N$, $\aveTau$ can be disregarded in Eq.\ (\ref{eq:aveNaLargeN}), simplifying to a power law $\aveNa \propto N ^ { \eta } $,      
\begin{equation} \label{eq:aveNaPowerLaw}                                  
    \lim_{N\to\infty} \aveNa \propto N^{ \beta_S \cdot (2 - \alpha_S) - \beta_D \cdot (2 - \alpha_D)} \textrm{ , for $ 1 < \alpha_S, \alpha_D < 2 $.}  
\end{equation}        

\begin{figure}[!htb]  
    \centering
    \includegraphics[width=\columnwidth]{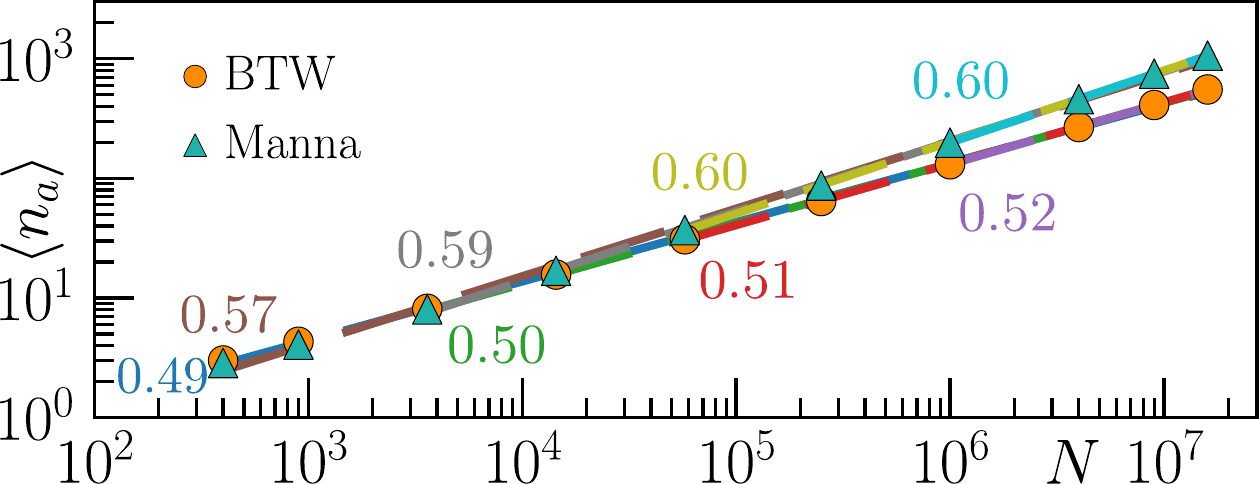}
    \caption{Same as in Fig.~\ref{fig:if-allometry}, but for the BTW (orange symbols) and Manna (cyan symbols) SOC models. From a power-law fitting $\aveNa \propto N^{\eta}$ (coloured dashed lines), we obtain sublinear allometric scalings with $\eta = 0.52 \pm 0.04$ (purple dashed line) for the BTW model and $\eta = 0.60 \pm 0.04$ (cyan dashed line) for the Manna model, considering $ L = \sqrt{N} \in [10^3 , 4 \cdot 10^3 ] $. Averages are performed over $10^7$ avalanches for each $N$, after reaching the steady-state.}
    \label{fig:soc-allometry}
\end{figure}

\begin{figure*}[!htb]             
    \centering
    \includegraphics[width=\textwidth]{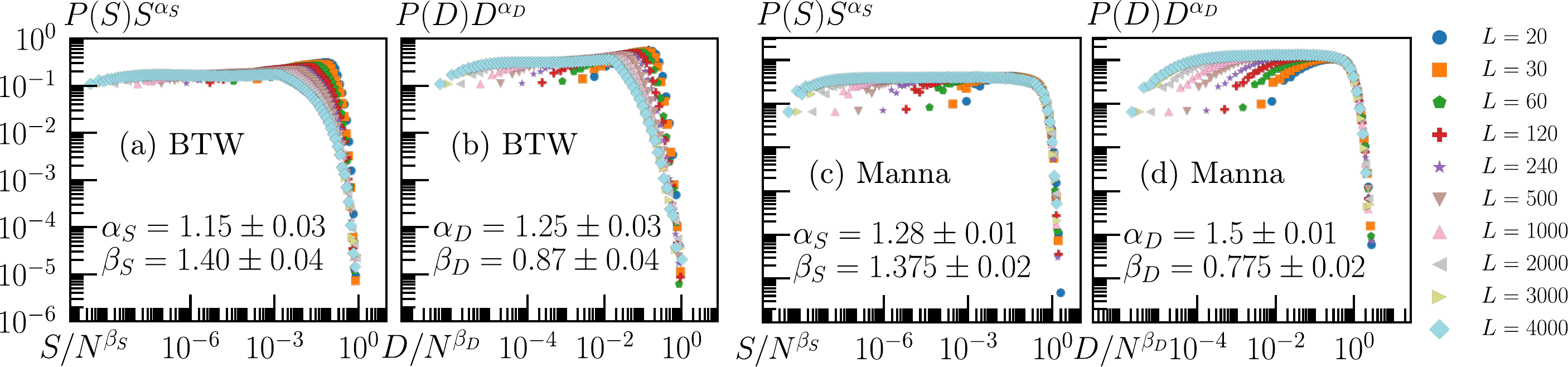}     
    \caption{Same as in Fig.\ \ref{fig:if-avalanches}, but for the BTW (a-b) and Manna (c-d) SOC models.}
    \label{fig:soc-avalanches}         
\end{figure*}        

Recalling that isometric scaling corresponds to $\eta = 1$, Eq.\ (\ref{eq:aveNaPowerLaw}) indicates that allometric scaling is generally expected. We can now compare this prediction with the exponents reported in Fig.\ \ref{fig:if-allometry}. First, notice that, since $\alpha_D > 2$ in the IF model (Figs.\ \ref{fig:if-avalanches}(b-d)), $\lim_{N\to\infty} \aveT$ is expected to converge to a constant, as previously mentioned. Therefore, Eq.\ (\ref{eq:aveNaPowerLaw}) modifies into    
\begin{equation} \label{eq:aveNaPowerLawModified}                                 
    \lim_{N\to\infty} \aveNa \propto N^{ \beta_S \cdot (2 - \alpha_S) } \textrm{ , for $ 1 < \alpha_S < 2 $, $\alpha_D > 2$.}  
\end{equation}        
Using the values $\{ \alpha_S , \beta_S \}$ estimated for the case $\pin = 0 \%$ (Fig.\ \ref{fig:if-avalanches}(a)), Eq.\ (\ref{eq:aveNaPowerLawModified}) predicts an allometric scaling exponent of $\eta = 0.48\pm0.03$, consistent with the fitting exponent of Fig.\ \ref{fig:if-allometry}. Likewise, for $\pin = 20 \%$ (Fig.\ \ref{fig:if-avalanches}(c)) Eq.\ (\ref{eq:aveNaPowerLawModified}) predicts $\eta = 0.38\pm0.03$, again consistent with the corresponding fitting exponent of Fig.\ \ref{fig:if-allometry}.             

\subsection{Sublinear allometry in self-organized criticality models}   

Finally, in order to verify that Eq.\ (\ref{eq:aveNaLargeN}) generally holds for critical avalanching systems, we consider two models for SOC, the BTW and Manna models, implemented on a square grid with $L^{2}=N$ sites and open boundaries \cite{bak_self-organized_1987,bak_self-organized_1988,manna_two-state_1991}. The main difference between the models is that the BTW sandpile has deterministic toppling dynamics, whereas in the Manna model the dynamics is stochastic. External driving is applied between avalanches by adding a single grain of sand at a randomly selected site at each timestep (more details of the implementation in Section \ref{section:Methods}).     

In Figs.\ \ref{fig:soc-allometry} and \ref{fig:soc-avalanches} we show the numerical results for $\aveNa$ as well as the data collapse of the distributions $P(S)$ and $P(D)$, for several $N \in [ 4 \cdot 10^2, 1.6 \cdot 10^7]$, for both SOC models. Specifically, in Fig.\ \ref{fig:soc-allometry} we observe sublinear allometric scaling for $\aveNa$, scaling as $\aveNa \propto N^{\eta}$, with $ \eta = 0.52 \pm 0.04 $ for the BTW model and $ \eta = 0.60 \pm 0.04 $ for the Manna model, considering $L = \sqrt{N} \in [ 10^3 , 4 \cdot 10^3 ]$ (purple and cyan dashed lines).     
The fitting of the average quiet time $\aveTau = aN^{b} + c$ in Fig.\ \ref{fig:aveTauNew}(c-d) gives $c=1.24\pm0.01$ for the BTW model and $c=0.39\pm0.01$ for the Manna model, with $b < 0$ for both cases, indicating that $\aveTau$ initially decays with $N$ as a power law, and eventually tends to a constant, thus Eq.\ (\ref{eq:aveNaPowerLaw}) is again valid for the SOC models.  
Using the respective parameters reported in Fig.\ \ref{fig:soc-avalanches}, the scaling exponent in Eq.\ (\ref{eq:aveNaPowerLaw}) is estimated to be $\eta = 0.54 \pm 0.09 $ for the BTW model and $\eta = 0.60 \pm 0.05 $ for the Manna model, consistent with the respective power-law fit $\aveNa \propto N^{\eta}$.
In variants of the BTW and Manna models in which the energy is kept fixed, the average density of active sites or rate of activation $\aveR$ is widely recognized as an order parameter \cite{vespignani_absorbing-state_2000,dickman_critical_2001,cessac_lyapunov_2001,giometto_connecting_2012,girardi-schappo_griffiths_2016,lee_universality_2017,girardi-schappo_brain_2021}, with the control parameter being the total energy (i.e., the number of grains, fixed by the initial conditions). Since $\rho$ is directly related to the average $\aveNa = N \rho \Delta t$, we can also examine the allometric scaling behaviour in these fixed-energy model variants.   
Interestingly, the fitted value $\eta= 0.52 \pm 0.04$ measured in Fig.\ \ref{fig:soc-allometry} for the BTW model is smaller than the one measured in the literature for the fixed-energy variant, with a value of $\eta = 0.60(8)$ for system sizes $L \in [ 320, 1280 ]$ \cite{vespignani_absorbing-state_2000}. 
For the BTW model, the mean-field critical exponents are $\alpha_S^{\textrm{MF}} = 3/2$, $\beta_S^{\textrm{MF}} = 2$, $\alpha_D^{\textrm{MF}} = 2$ and $\beta_D^{\textrm{MF}} = 1$ \cite{chessa_mean-field_1998, zapperi_self-organized_1995}. Since we have the particular case $\alpha_D^{\textrm{MF}} = 2$, $\lim_{N\to\infty} \langle D \rangle \propto \ln{N}$ (see supplementary information), and Eq.\ (\ref{eq:aveNaLargeN}) is instead replaced by $\lim_{N\to\infty} \aveNa^{\textrm{MF}} \propto N^{ \beta_S^{\textrm{MF}} \cdot ( 2 - \alpha_S^{\textrm{MF}} ) } / \left( c_D \ln{N} + \langle \tau \rangle \right)$. Replacing the values of $\alpha_S^{\textrm{MF}}$ and $\beta_S^{\textrm{MF}}$, and assuming $\lim_{N\to\infty} \langle \tau \rangle \propto \textrm{constant}$ in the mean-field limit, we have $\lim_{N\to\infty} \aveNa^{\textrm{MF}} \propto N/\ln{N}$, different from the power-law scaling measured numerically in Fig.\ \ref{fig:soc-allometry}, as expected since the dimension $d = 2$ is still below the upper critical dimension $ d_c > 4 $ \cite{chessa_mean-field_1998}.      
On the other hand, the fitted exponent $\eta = 0.60 \pm 0.04$ measured in Fig.\ \ref{fig:soc-allometry} for the Manna model is consistent with the one measured in the literature for the fixed-energy variant of this model, with a value of $\eta = 0.60(9)$, considering system sizes $L \in [32, 1024]$ \cite{vespignani_absorbing-state_2000}.              

\subsection{Sublinear allometry in real cortical systems}      

Experimental evidence, both from indirect measurements of glucose utilization rates \cite{karbowski_thermodynamic_2009} and from energy conservation arguments \cite{herculano-houzel_scaling_2011,karbowski_scaling_2011,lennie_cost_2003, yu_evaluating_2017}, indicates that neuronal firing rates decrease with brain size. In mammals, the average firing rate $\rho$ of neurons in gray matter, estimated indirectly from measurements of glucose utilization rates, decreases with the brain size, scaling with gray matter volume as $\rho \propto V_G^{-0.15}$ \cite{karbowski_thermodynamic_2009}. More specifically, the total firing rate $\langle n_a \rangle$ should scale as $\langle n_a \rangle = N \rho \propto NV_G^{-0.15}$. If we denote $\gamma$ the scaling exponent of the gray matter volume $V_G$ with the number of cortical neurons $N$, $V_G \propto N^{\gamma}$, then $\langle n_a \rangle \propto N^{1-0.15 \gamma}$. We see that the condition to have sublinear allometry for $\aveNa$ with $N$ is only that $\gamma>0$, that is, the gray matter volume must increase with the number of cortical neurons.
Indeed, this is verified from experimental data in \cite{ventura-antunes_different_2013}, where the authors estimate $\gamma = 1.587$ and $\gamma = 0.918$ using data across rodent and primate species, respectively. This gives an estimate of $\aveNa \propto N^{0.762}$ across rodent species and $\aveNa \propto N^{0.862}$ for primates, qualitatively consistent with the “economy of scale” behaviour seen for $\langle n_a \rangle$ in the IF model.  
Furthermore, in \cite{herculano-houzel_scaling_2011,karbowski_scaling_2011}, the authors argued that a fixed energy budget per neuron imposes constraints on neuronal activity, so that neurons in larger brains are on average less active than in smaller brains. In \cite{lennie_cost_2003}, the author compares the energy cost of spiking in cortical neurons between humans and rats, explaining that spikes from human cortical neurons are more expensive due to a greater number of synapses and longer dendrites and axon collaterals. It is then concluded that, as a result, the human cortex must have significantly fewer active neurons than the smaller rat cortex. This is supported by the findings in \cite{yu_evaluating_2017}, where authors reported that the average cortical firing rate of excitatory neurons in the awake resting state is higher in rats, $\sim 4.3$Hz, compared to humans, $\sim 1.15$Hz.
The generic allometric behaviour of systems with critical avalanche dynamics disclosed in this paper might thus explain the observed allometry of brain activity.        
 
\section{Discussion}  

We have shown that the sublinear scaling of brain activity is not an incidental trait of neural circuits, but a robust manifestation of avalanche dynamics. By deriving scaling laws from universal avalanche statistics, we demonstrate that any system governed by critical avalanches must display an “economy of scale” in its activity. This result holds independently of microscopic details, and it is corroborated in both integrate-and-fire neuronal networks and canonical self-organized criticality (SOC) models. 
Strikingly, the exponents we predict align with empirical measurements across mammal species, bridging theory and biological reality. Both brain allometry and Kleiber’s metabolic law ultimately trace back to branching structures—in metabolism, the vascular network that distributes energy, and in avalanching systems, the spread of activity across a network at criticality. In the mean-field limit, avalanche propagation maps exactly onto a critical branching process, yielding the experimentally measured avalanche size exponent $\tau = 3/2$ \cite{zapperi_self-organized_1995}. 
By identifying avalanche dynamics as the origin of neuronal firing allometry, our work highlights critical avalanches as the universal dynamical mechanism that constrains efficiency in large neural systems. This perspective deepens our understanding of why larger brains can function with disproportionately lower firing activity, and points to avalanche criticality as a general organizing principle for scalable, resilient, and efficient complex systems.   

\section{Methods} \label{section:Methods} 

\subsection{IF model}  \label{subsection-implementation-if}

We implement an integrate-and-fire (IF) model on a scale-free, directed network considering both short- and log-term plasticity \cite{michiels_van_kessenich_critical_2018}, as well as a refractory time, an interval of one timestep during which neurons remain inactive immediately after firing. We consider systems with a different number of neurons $N$ randomly placed in a cube of side $L$, keeping the density of neurons $N / L^{3} = 0.016$ fixed. We consider a percentage of inhibitory neurons $ \pin$. The outgoing degree $k$ of each neuron follows a power-law, $P(k) \propto k^{-2}$, with $k \in [2, 100]$, and the probability for two neurons to be connected decays exponentially with their euclidean distance $r$, $P(r) \propto e^{ - r / r_{0} }$, where $r_{0} = 5$ is a characteristic length. Each neuron $i$ is characterized by its potential $v_i$, where the resting state is $v_{i} = 0$. Neuron $i$ will fire every time its potential is larger than a threshold, $v_i \geq v_c \equiv 1$, transmitting its signal to all its post-synaptic neurons, 
\begin{align}	    
\label{eq:synaptic-transmition} 	v_j (t+1) &= v_j(t) \pm v_i(t) u_i(t) g_{ij}  \textrm{ ,} \\    
\label{eq:STP}	u_i (t+1) &= u_i(t) \cdot (1 - \delta u) \textrm{ ,} \\    
\label{eq:v-reset} 	v_i (t+1) &= 0 \textrm{ ,}    
\end{align}     
where $+$ and $-$ stands for excitatory and inhibitory pre-synaptic neuron, respectively, $u_i(t) \in [0,1]$ and $g_{ij} \in [10^{-5},1]$ represent the amount of neurotransmitters and the synaptic strengths, respectively modelling the short- and long-term plasticity, and $\delta u=0.05$ accounts for the fractional amount of neurotransmitter released at each neuronal firing. To keep activity ongoing, a small external input of $\delta v = 0.1 v_{c}$ is added to a random neuron at each timestep between avalanches. At the end of each avalanche, the amount of neurotransmitters is recovered by a certain amount $\durec$, $u_i(t) \rightarrow u_i(t) + \durec$. By carefully tuning the parameter $\durec$ as a function of the system size $N$ and fraction $\pin$, the system can exhibit avalanches of size $S \geq 1$ and duration $D \geq 1$ that are power-law distributed, indicating that it is operating at a critical point.             
For all simulations, we average quantities over $2000$ different network configurations, and over $2 \cdot 10^{4}$ avalanches for each configuration. Before making any measurements, we let the dynamics evolve for either $10^{4}$ avalanches or until a synaptic strength $g_{ij}$ first reaches a minimum value $ g_{\textrm{min}} = 10^{-5}$, in which case we set that strength to $g_{ij} = g_{\textrm{min}}$, in order to shape the distribution of synaptic strengths $g_{ij}$ following the rules of Hebbian plasticity \cite{hebb_organization_1949}, as described in \cite{michiels_van_kessenich_critical_2018}.        

\subsection{BTW model} \label{subsection-implementation-btw} 

The 2D Bak–Tang–Wiesenfeld (BTW) sandpile model consists of a square grid with $L \times L$ sites, with each site $(x \in [1,L] , y \in [1,L])$ defined by its number of grains or height $h_{x,y} \geq 0$, with open boundaries $h_{0,y},h_{x,0},h_{L+1,y},h_{x,L+1}=0$. Whenever the height at some site $h_{x,y}$ reaches a given threshold $h_c \equiv 4$, i.e. $ h_{x,y} \geq h_c $, the site $(x,y)$ topples (i.e. becomes active), $h_{x,y} \rightarrow h_{x,y} - 4 $, and one grain is added to each of its four nearest neighbours, $ h_{x\pm1, y} \rightarrow h_{x\pm1, y} + 1 $ and $ h_{x, y\pm1} \rightarrow h_{x, y\pm1} + 1$. If any of these four neighbours is at a boundary, the grain exits the system. This mechanism generates avalanches of size $S \geq 1$, corresponding to the number of topplings that occur during the avalanche, and duration $D \geq 1$, corresponding to the number of timesteps the avalanche lasts. When an avalanche ends, a grain is added to a random site $ h_{x,y} \rightarrow h_{x,y} + 1$ each timestep until a site reaches the threshold, and avalanche activity is resumed. This time interval with no activity between avalanches is the quiet time $\tau \geq 0$.  
The dynamics is implemented with parallel updating \cite{vespignani_absorbing-state_2000}, i.e. all active sites $h_{x,y} \geq h_c$ topple in the same timestep.  
For all simulations regarding the SOC model, we average quantities by generating over $10^7$ avalanches, after reaching the steady state, considering sizes of $ L \in [20,4000] $.   

To ensure that the system has indeed reached the steady state, we initially monitor the average height $\overline{h}(t) = \frac{1}{N} \sum_{x,y}^{L} h_{x,y}(t)$ at a given timestep $t$ and compute the relative difference $ \varepsilon $ between the current average height and the one at a previous timestep $t - t_p$, $ \varepsilon = | \overline{h}(t) - \overline{h}(t-t_p) | / \overline{h}(t-t_p) $, with $t_p = 3L$ increasing with the system size to account for the slow down in the change over time of $\overline{h}$ for larger systems, and declare that the system has reached the steady state after observing that the difference is consistently below a certain threshold $ \varepsilon \leq 10^{-4} $. 
 
\subsection{Manna model}  \label{subsection-implementation-manna}     
 
The implementation of the 2D Manna model is analogous to the BTW model except for the toppling rule, which is now stochastic. In the Manna model, the threshold is defined as $h_c \equiv 2$. Whenever a site $(x,y)$ reaches this threshold $h_{x,y} \geq h_c$, the site topples, $h_{x,y} \rightarrow h_{x,y} - 2$, with the two toppled grains added to two randomly selected nearest neighbour sites $( x' \in \{ x \pm 1 \} , y )$ or $( x , y' \in \{ y \pm 1 \} )$, chosen independently (two grains can fall onto the same neighbouring site with probability $1/4$). The dynamics is also implemented using parallel updating. The imposed boundary conditions and the monitoring procedure to reach the steady state are identical to the BTW model.     
 
\appendix 
   
\section*{Acknowledgments}     
LdA would like to acknowledge the support by \#NEXTGENERATIONEU (NGEU) funded by the Ministry of University and Research (MUR), National Recovery and Resilience Plan (NRRP), project MNESYS (PE0000006)—A multiscale integrated approach to the study of the nervous system in health and disease (DN. 1553 11.10.2022). H.J.H. thanks FUNCAP and INCT-SC for financial support. J.S.A.J. thanks the Brazilian agencies CNPq, CAPES and FUNCAP for financial support. SZ acknowledges support from CAPES (PrInt, process n. 88887.937759{/}2024-00).

\bibliography{ references-main }
  
\end{document}